\documentclass[%
 reprint,
 amsmath,amssymb,
 aps,
]{revtex4-2}

\usepackage{graphicx}
\usepackage{dcolumn}
\usepackage{bm}
\usepackage{subcaption}
\usepackage[version=4]{mhchem}
\usepackage{siunitx}
\DeclareSIUnit\Molar{\textsc{m}}

\begin{document}

\preprint{APS/123-QED}

\title{\textit Stochastic Kinetic Study of Protein Aggregation and Molecular Crowding Effects of $A\beta40$ and $A\beta42$}

\author{John Bridstrup}%
\affiliation{%
 Department of Physics, Drexel University, Philadelphia, PA 19104, USA
}%
\author{Jian-Min Yuan}%
\affiliation{%
  Department of Physics, Drexel University, Philadelphia, PA 19104, USA
}%
\author{John S. Schreck}
\affiliation{National Center for Atmospheric Research (NCAR), Computational and Information Systems Lab, Boulder, CO, USA
}%

\date{\today}

\begin{abstract}
Two isoforms of $\beta-$amyloid peptides, A$\beta$40 and A$\beta$42, differ from each other only in the last two amino acids, IA, at the end of A$\beta$42. They, however, differ significantly in their ability in inducing Alzheimer's disease (AD). The rate curves of fibril growth of A$\beta$40 and A$\beta$42 and the effects of molecular crowding have been measured in \textit{in  vitro} experiments. These experimental curves, on the other hand, have been fitted in terms of rate constants for elementary reaction steps using rate equation approaches. Several sets of such rate parameters have been reported in the literature. Employing a recently developed stochastic kinetic method, implemented in a browser-based simulator, \textit{popsim}, we study to reveal the differences in the kinetic behaviors implied by these sets of rate parameters. In particular, the stochastic method is used to distinguish the kinetic behaviors between A$\beta$40 and A$\beta$42 isoforms. As a result, we make general comments on the usefulness of these sets of rate parameters.

\end{abstract}

\maketitle


\section{\label{sec:level1}Introduction}

		Protein aggregation into amyloids is often  the culprit for neurodegenerative diseases such as Alzheimer’s disease (AD) and Parkinson’s disease. There are about 20 such cases \cite{dobson2003protein,knowles-09}.  In the earlier version of the amyloid hypothesis of Alzheimer’s disease \cite{hardy02amylhypothesis}, it was assumed that the self-assembly of $\beta$-amyloid peptides, such as $A\beta40$, $A\beta42$, into amyloids and plaques was the main cause of the disease, but later studies show that the most neurotoxic species are intermediate soluble oligomers and the amyloid plaque itself may even have some protective functions of the neurons \cite{urbanc17}.  But the specific sizes of the neurotoxic oligomers are still not unambiguously identified. $A\beta42$ is considered to be a bio-marker for AD, for it was discovered that AD patients tend to have a lower concentration of $A\beta42$ in their cerebrospinal fluids (SPF), or even better, lower A$\beta$42/A$\beta$40 ratio \cite{lewczuk}.  In fact, recently, US FDA approved (on May 4, 2022) the marketing of first diagnostic test for early detection of amyloid plaques associated with AD, Lumipulse G $\beta$-amyloid ratio (1-42/1-40) test, using SPF for 55 and older patients.  The two isoforms of $\beta$-amyloid peptides, $A\beta40$ and $A\beta42$, differ in the extra amino acids, I and A, in the amino acid sequence of $A\beta42$. The amino acid sequence for $A\beta42$ is: DAEFRHDSGYEVHHQKLVFFAEDVGSNKGAIIGLMVGGVVIA. Since I and A are hydrophobic amino acids, it is argued that this higher overall hydrophobicity of $A\beta42$  tends to yield higher aggregation rates, compared to $A\beta40$.  
		Experiments, using copper and hydrogen peroxide induced cross-linking of unmodified proteins, the PICUP method, have indeed shown that polymer formation advances at a faster rate for $A\beta42$ than $A\beta40$ \cite{urbanc17}. The results are also supported by Molecular dynamics studies using all-atom molecular dynamics (MD) \cite{mittal2015macromolecular} discrete molecular dynamics (DMD) \cite{urbanc17}.  
		The cellular environments of proteins are crowded with other biomolecules, such DNA,  RNA, lipids, and other proteins inside cellular compartments.  This crowdedness due to other macromolecules, called macromolecular crowding \cite{ellis,Minton98,Minton-14,ping06,schreck17}, and also confinement \cite{ping2003effects}, affect the reaction rates, properties, and functions of specific proteins of interest.  More specifically, for the problem of interest here, macromolecular crowding \cite{Schreck20} or confinement in the extracellular space in the brain \cite{nicholson,nicholson20} may affect the aggregation rates of $\beta$-amyloid peptides and affect the rates differently for $A\beta40$ and $A\beta42$.  

		Experimental data on the rates of fibril formation have often been fitted using rate equations \cite{meisl16,meisl2016molecular,Schreck20}, where mass-action laws are employed for the elementary chemical reactions that may be involved. Rich information have been accumulated by these rate equation approaches. along with our understanding of the amyloid formation problem. In particular, several sets of rate parameters  for A$\beta$40 and A$\beta$42 have been reported through such fitting \cite{meisl14,cohen2013proliferation,Schreck20}.
		
		More recently it has been found that in the case of biochemical reactions taking place in very small volumes, stochastic kinetic approaches are more appropriate \cite{allen,tiwari,bridstrup21,shen21stochastic}. This is because, the concept of concentrations may not be well-defined when the fluctuations for the numbers of chemical species are too large in a tiny volume. In fact, fluctuations for many dynamical variables all can become large in these cases.  The focus of the present work is on a stochastic kinetic approach that we have recently developed \cite{bridstrup21} and apply it to study the kinetic behaviors implied by the several sets of rate parameters for A$\beta$40 and A$\beta$42 mentioned above. The merit of such a stochastic kinetic study partly lies in the fact that during the growth period of the aggregates, the populations of dimers, trimers. tetramers, pentamers,... all of their populations start from zero, go thru small numbers and eventually reach their steady-state numbers. such a process is more properly studied using a stochastic approach instead of a rate equation one, as discussed above.   

		The organization of the article is as follows: We introduce the theoretical background of the elementary reaction mechanisms and the effects of molecular crowding considered in both rate-equation and stochastic kinetic approaches. This is followed by a brief presentation of a stochastic kinetic approach (based on the rate-equation formulation) and a browser-based stochastic numerical simulator that we have developed in Section III. In Section IV, we present and discuss the results of our stochastic kinetic calculations on five sets of rate parameters for $A\beta40$ and $A\beta42$, along with a study of the effects of molecular crowding on these two isoforms. We conclude by a comment section.

\section{Theoretical Background and Simulation Schemes}
In this section we introduce the reader to \textit{popsim}, a stochastic simulation application developed by the authors. We also give a brief overview of the models and methods implemented. \textit{popsim} is a browser-based simulation package, written in Typescript, which generates stochastic pathways based on the reaction mechanisms of several well-known models for protein aggregation using the Gillespie method \cite{gillespie07,gillespie92,gillespie77}. It gives the user several visualization tools for the data, as well as the option to download data for more in depth exploration. The application is user-friendly, and has use in both in-depth research studies as well as teaching and demonstration settings.

\textit{popsim} has several built-in models of protein aggregation, each composed of a combination of several individual reaction mechanisms. Each mechanism describes a single reaction or a single "type" of reaction which may occur within the system volume. In this section, we will briefly describe the various reaction mechanisms in terms of their chemical reactions and show which combinations are included in the various models. Some of the reaction details can be found also in \cite{bridstrup16,bridstrup21,Schreck20} and their connections to the stochastic kinetic model pointed out below as well as in \cite{bridstrup21}.
\subsection{Basic reaction mechanisms}
\subsubsection{Monomer addition and subtraction}
Monomer addition, and its reverse, monomer subtraction, is the reaction mechanism through which protein aggregates grow and shrink by gaining or losing one monomer at a time. It is described by the chemical reaction
\begin{equation*}
\ce{ M_1 + M_r <=>[a][b] M_{r+1} }, 
\end{equation*}
where $M_r$ is an aggregate composed of $r$ monomers (an $r$-mer) and $a$ and $b$ are the rate constants for the forward reaction (addition) and reverse reaction (subtraction), respectively. These two mechanisms are, by themselves, a simple model of aggregation also known as the Becker-Doring model of nucleation \cite{BD,Wattis}. The rate equations describing these reactions are
\begin{align*}
    \frac{dc_r}{dt} &= ac_1(c_{r-1} - c_r) + b(c_{r+1}-c_r)\\
    \frac{dc_1}{dt} &= -\sum_{r=n_c}r\frac{dc_r}{dt},
\end{align*}
where $c_r$ is the concentration of $r$-mers and the associated propensity functions, which are weighting factors for specific reactions to be used in the Gillespie stochastic simulation algorithm \cite{gillespie07,bridstrup21}, are given by
\begin{equation*}
    W_{r+} = a'N_1N_r,
\end{equation*}
for addition and
\begin{equation*}
    W_{r-} = b'N_r,
\end{equation*}
for subtraction, where $N_r$ is the population of $r$-mers and the primes signify that we are using the stochastic rate constants, defined below.
\subsubsection{Coagulation and fragmentation}
In real systems, aggregates may also merge with (coagulate) or break from (fragment) each other. This is described by the chemical reaction
\begin{equation*}
\ce{ M_r + M_s <=>[k_+][k_-] M_{r+s} }, 
\end{equation*}
where $k_+$ and $k_-$ are the rate constants for coagulation and fragmentation, respectively. The set of rate equations describing the change in concentration of $r$-mers due to these reactions is given by
\begin{align*}
    \frac{dc_r}{dt} &= k_+\Bigg[\frac{1}{2}\sum_{s=n_c}^{r-n_c}c_sc_{r-s} - \sum_{s=n_c}^{\infty}[1 + \delta_{rs}]c_rc_s\Bigg]\\
    &+k_-\Bigg[\sum_{s=2}^{\infty}[1 + \delta_{rs}]c_{r+s} - (r - 3)c_r\Bigg],
\end{align*}
where $\delta_{rs}$ is the Kronecker delta function, and the propensity functions are given by
\begin{equation*}
    W_{rs+} = k_+'N_r(N_s - \delta_{rs})
\end{equation*}
for an $r$- and $s$-mer becoming an $(r+s)$-mer and
\begin{equation*}
    W_{rs-} = k_-'N_{r+s}.
\end{equation*}
The $\delta_{rs}$ is to account for the fact that two of the same sized polymers may combine. Again primes denote the stochastic rate constants defined later. 
\subsubsection{Primary Nucleation}
Primary nucleation process, by which a \textit{critical nucleus} is formed from $n_c$ monomers, is described by the chemical reaction
\begin{equation*}
    \ce{ n_cM_1 <=>[k_n][k_n^*] M_{n_c} },
\end{equation*}
where $k_n$ is the primary nucleation rate constant and $k_n^*$ is a model-dependant composite rate constant for the dissociation of sub-critical polymers. The rate equation describing how the concentration of \textit{critical nuclei} changes is
\begin{equation*}
    \frac{dc_{n_c}}{dt} = k_nc_1^{n_c}.
\end{equation*}
Thus here we neglect the reverse dissociation reaction, which is often dominated by the forward reaction.   The propensity function for this reaction is
\begin{equation*}
    W_n = k_n'\prod_{s=0}^{n_c-1} (N_1-s).
\end{equation*}
When primary nucleation is included in a model, it is assumed that no polymers smaller than $n_c$ exist, as such the rate constant $k_n^*$ depends on all other mechanisms which would produce a sub-critical polymer.
\subsubsection{Secondary nucleation}
Secondary nucleation, or heterogeneous nucleation, is an auto-catalytic process where new aggregates are formed on the surface of existing aggregates. Secondary nucleation can be modeled by a two-step process \cite{meisl14,Schreck20} given by the chemical reactions
\begin{align*}
\ce{ n_2M_1 + M_S &<=>[k_f][k_b] (M_{n_2} + M_S)_{bound} \\
&->[\Bar{k_2}] M_{n_2} + M_S }, 
\end{align*}
where $M_S$ is a measure of the available surface active sites to which monomers can bind and $k_f$, $k_b$ and $\bar{k_2}$ are the rate constants for surface attachment and detachment of monomers, and the production of new secondary nuclei, respectively. An in-depth discussion of this mechanism can be found in \cite{meisl14,Schreck20}, but it is worth noting that a limiting case of this mechanism is given by the catalytic chemical reaction
\begin{equation*}
\ce{ n_2M_1 + M_S ->[k_2] M_{n_2} + M_S },
\end{equation*}
otherwise known as one-step secondary nucleation. In one-step nucleation, the three rate constants $k_f$, $k_b$ and $\bar{k_2}$ reduce to a single rate constant for the production of nuclei, $k_2$. As of this writing, \textit{popsim} includes only one-step secondary nucleation, whose propensity function is given by
\begin{equation*}
    W_2 = k_2'N_S\prod_{s=0}^{n_2-1}(N_1-s),
\end{equation*}
where $N_S$ is the number of surface active sites available to monomers.
\subsection{Available models}
\textit{popsim} includes three basic models of protein aggregation. They are combinations of mechanisms described in the previous section. None of the default models include the effects of crowders or secondary nucleation, but the user may select versions of the model which do.
\subsubsection{Oosawa model}
The Oosawa model \cite{oosawa,bridstrup16} is a combination of primary nucleation, monomer addition and monomer subtraction. It is the simplest model available, but describes very well many aggregating polymers such as actin \cite{bridstrup16,Schreck20}.
\subsubsection{Smoluchowski model}
The generalized Smoluchowski model combines coagulation and fragmentation with the Oosawa model. Due to the additional mechanisms, there are a wider variety of systems for which this model is useful \cite{Schreck13,schreck17,Schreck20}. Additionally, limiting cases of this model (ie. without fragmentation, or without coagulation) can be investigated by simply setting the appropriate rate constants to $0$.
\subsubsection{Smoluchowki with Becker-Doring nucleation}
This is the most complex of the three models. It is a combination of simple monomer addition and subtraction up to a certain size ($r<n_c$), above which aggregates behave according to the generalized Smoluchowski model ($r\geq n_c$). Effectively, this model replaces Oosawa primary nucleation with a more accurate model of nucleation \cite{Wattis,wattis1999becker}. 
\subsection{Effects of macromolecular crowders}
The effects of crowding molecules are included through use of formulas found by applying the scaled particle theory (SPT) of fluid mixtures \cite{SPT,hall02,hall04,Ferrone02,Minton-14,bridstrup16,Schreck20}. We give only the resulting equations in this section. Interested readers may refer to the references for a detailed derivation. Effectively, crowding plays prominent role  only for growth processes (nucleation, coagulation, etc..). For coagulation, the rate constant is found to be
\begin{equation*}
    k_+ = \frac{\gamma}{\alpha}k_+^0,
\end{equation*}
where $k_+^0$ is the rate constant without crowders and $\gamma$ and $\alpha$ are calculated from SPT. The rate constant for monomer addition is a specific case of the previous equation and is identical in form. The primary nucleation rate constant is
\begin{equation*}
    k_n = \bigg(\frac{\gamma}{\alpha}\bigg)^{n_c-1}k_n^0
\end{equation*}
and one-step secondary nucleation
\begin{equation*}
    k_2 = \gamma^{n_2}\Gamma k_2^0,
\end{equation*}
where superscript $0$ again denotes the crowderless rate constant value and $\Gamma$ is another factor from SPT \cite{Ferrone02}. For details on how to calculate the SPT parameters, the reader may refer to the references \cite{hall02,hall04,Schreck20}. \textit{popsim} assumes spherical crowders, spherical monomers and sphero-cylindrical aggregates. The variables required for simulation are:
\begin{enumerate}
    \item monomer radius, $r_1$
    \item crowder radius, $r_c$
    \item spherocylindrical radius, $r_{sc}$, and
    \item crowder volume fraction, $\phi$.
\end{enumerate}
Each model in \textit{popsim} has the option to include the effects of crowders, making it a very powerful tool for exploring the dynamics of proteins within inherently crowded living cells.

\section{Gillespie stochastic simulation algorithm}
\textit{popsim} implements the Gillespie stochastic simulation algorithm (SSA), which is a method for generating statistically correct trajectories of a system of molecules undergoing discrete reaction steps. It is particularly useful for investigating the behavior of small numbers of reacting particles (\textit{ie}. small system volumes), for which number fluctuations are comparable to the total number of reactants and continuous rate equations are inaccurate in describing the reaction processes. The algorithm is given in \cite{bridstrup21} and included here for completeness: 
\begin{enumerate}
    \item Set the initial species populations and set $t=0$.
    \item Calculate the transition rate or the propensity function, $r_i$, for each of the $M$ possible reactions.
    \item Set the total transition rate $Q = \sum_{i=1}^{M}r_i$.
    \item Generate two uniform random, (0,1),  numbers, $u_1$ and $u_2$.
    \item Set $\Delta t = \frac{1}{Q}ln(\frac{1}{u_1})$.
    \item Find $\mu \in [1,..,M]$ such that \\ \\
    $\sum_{i=1}^{\mu -1}r_i < u_2Q \leq \sum_{i=1}^{\mu}r_i$.
    \item Set $t = t + \Delta t$ and update molecular species based on reaction $\mu$.
    \item Return to step 2 and repeat until an end condition is met.
\end{enumerate}
\subsection{Calculating stochastic rate constants}
In the previous sections, we have discussed chemical species in terms of their bulk concentration in solution. Since \textit{popsim} deals with discrete populations of species, it is necessary to relate the bulk rate constants, which are volume independent, to the volume-dependent stochastic rate constants. As an example, the bulk rate of change in the concentration of nuclei due to primary nucleation is given by
\begin{equation*}
\frac{dc_{n_c}}{dt} = k_nc_1^{n_c},
\end{equation*}
where all $c$ are in units of molar concentration. Using the relationship
\begin{equation*}
    c_i = \frac{N_i}{N_AV},
\end{equation*}
where $N_i$ is the number of species of size $i$, $V$ is the system volume and $N_A$ is Avogadro's number, the rate of change of the number of nuclei due to primary nucleation is found to be
\begin{equation*}
    \frac{dN_{n_c}}{dt} = \frac{k_n}{(N_AV)^{n_c-1}}N_1^{n_c}.
\end{equation*}
Again, making use of the relationship between concentration and volume, the stochastic rate constant is found to be
\begin{equation*}
    k_n' = \bigg(\frac{c_0}{N_0}\bigg)^{n_c-1}k_n,
\end{equation*}
where $N_0$ and $c_0$ are the initial number and bulk concentration of monomers, respectively. The other stochastic rate constants (') are given, relative to their bulk counterparts, as follows:
\begin{align*}
    a' &= \bigg(\frac{c_0}{N_0}\bigg)a\\
    k_+' &= \bigg(\frac{c_0}{N_0}\bigg)k_+\\
    k_f' &= \bigg(\frac{c_0}{N_0}\bigg)k_f\\
    k_2' &= \bigg(\frac{c_0}{N_0}\bigg)^{n_2}k_2\\
    b' &= b\\
    k_-' &= k_- \\
    k_b' &= k_b \\
    \bar{k_2'} &= \bar{k_2}.
\end{align*}
\textit{popsim} makes these calculations automatically, thus the user need only have bulk system parameters, which are much more readily available and interpreted.
\subsection{Data visualization in \textit{popsim}}
Multiple ways to visualize data are provided in \textit{popsim}. Time evolutions of the average aggregate mass, M, average length of the filamentous aggregates, $<L>$, and average number of fibrils, P can be visualized. The averages are taken over many individual runs.  On the other hand, M and L values of individual runs as functions of time are also available to visualize. Furthermore, the standard deviations of M, L, and P as functions of time can also be plotted. In \textit{popsim} you can also plot the histogram of aggregate size distribution as it evolves as a function of time. Finally, the reaction frequency or propensity plot to rank each reaction mechanism can be plotted at various reaction times.

\section{Results and discussion}
In this section, using Gillespie's stochastic method we study the kinetic behaviors of five sets of rate parameters reported in the literature. These sets of rate parameters, for convenience, have been labeled as Ab40, Ab42-JSS, Ab42-Cohen, Ab40-Meisl, and Ab42-Meisl,  are listed in Table I. The first three sets of rate parameters, Ab40, Ab42-JSS, and Ab42-Cohen, are given in Ref. \cite{Schreck20} and the last two sets are listed in Ref. \cite{meisl14}. We present and discuss the results that we obtain by carrying out simulations for these sets of rate parameters using the stochastic kinetic method, i.e., the \textit{popsim} simulator, presented in the earlier sections.

\subsection{Ab40 versus Ab42-JSS}

\begin{figure}[ht]
    \centering
    \includegraphics[width=\columnwidth]{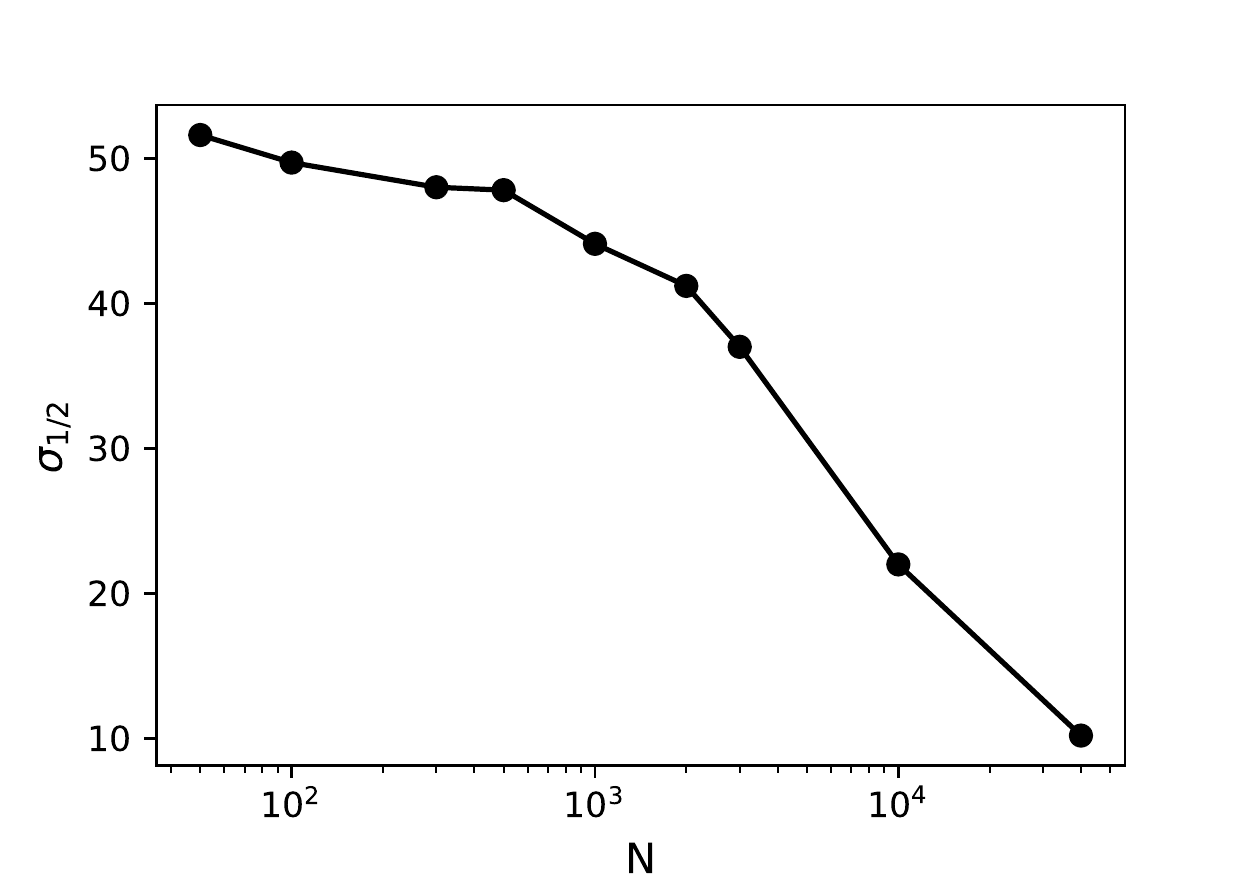}
    \caption{The standard deviation of the aggregate mass at halftime, $\sigma_{1/2}$, as a function of N for the Ab40 parameter set.}
    \label{fig1}
\end{figure}

\begin{figure*}[ht]
    \centering
    \includegraphics[width=2\columnwidth]{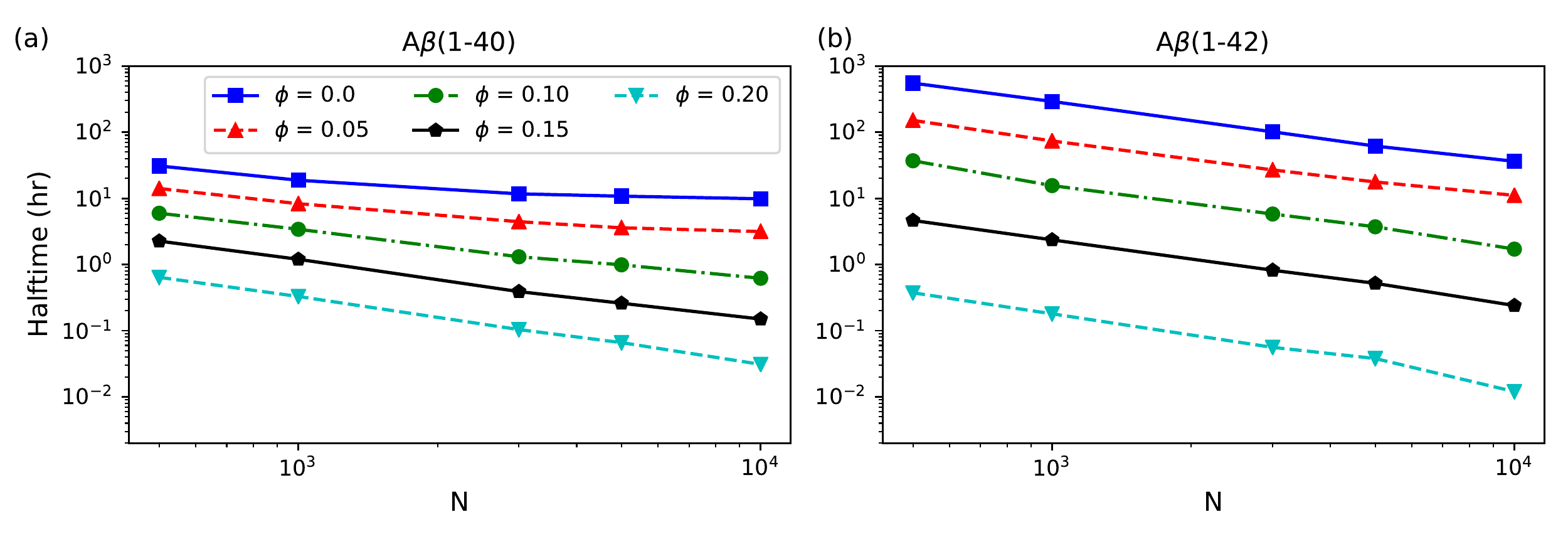}
    \caption{The halftime as a function of N and $\phi$ for (a) the Ab40 set (b) the Ab42-JSS set.}
    \label{fig2}
\end{figure*}

In this subsection, we consider the differences in kinetic behaviors represented by the two sets of reaction parameters: Ab40 and Ab42-JSS. In general we examine the average behaviors of many individual stochastic runs in \textit{popsim} to determine, for instance, whether a system has reached steady-state or equilibrium.  As mentioned in previous section, individual runs can behave quite different from the average. A measure of such fluctuations is standard deviation of a certain quantity. One important indicator of the kinetic behavior is the halftime, $t_{1/2}$, which is here defined by the time when half of the protein is found to be in aggregated form. To begin with, we study the fluctuations of the aggregate mass, M. In \textit{popsim} we can calculate the standard deviation of M as a function of time and it turns out the maximal standard deviation usually occurs at halftime.  We calculate the standard deviation at halftime relative to the maximal steady-state M value and call it $\sigma_{1/2}$, which is plotted for the Ab40 set in Fig. 1. The figure shows that for Ab40 set, the relative standard deviation increases to about 50$\%$ as N decreases. High fluctuations in M and other quantities are common for all the parameter sets investigated here.  Next, focusing on the volume effect, we have calculated $t_{1/2}$ as a function of N, the initial total number of monomers of an amyloid peptide, assuming initially the only amyloid species present are the monomers. If we keep the concentration constant among different experiments, smaller N thus represents smaller reaction volume.   Furthermore, we investigate how $t_{1/2}$ is affected by the presence of molecular crowders. We quantify the amount of molecular crowders present in the solution by, $\phi$, the ratio of the volume that they occupied to the total volume of the solution.
In Fig. 2(a) and Fig. 2(b), we present $t_{1/2}$ as a function of N and $\phi$.
From these two figures the general trend is clear, as N or volume increases, $t_{1/2}$ decreases, that is, average aggregation rate speeds up. A statistical interpretation of this is that $t_{1/2}$ is bounded by (0, $\infty$), thus increased importance of fluctuations will tend to push $t_{1/2}$ to the right at small enough N.  With $\phi$, as the amount of crowders increases $t_{1/2}$ decreases
due to entropy effect \cite{schreck17,Schreck20}.  However, comparison between the kinetic behaviors of A$\beta$40
and A$\beta$42, based on these two sets of rate parameters leads to an unexpected result that it seems that A$\beta$40 has a higher aggregation rates than A$\beta$42.  It seems to lead something that is contrary to what we would have expected, that is, A$\beta$42 had a higher aggregation rates than A$\beta$40. This could mean we would not be in favor of choosing Ab42-JSS as our set of parameters, until we examine other aspects of the  reaction kinetics.  Using \textit{popsim} we can easily obtain the time evolution of the average length of the fibrils and that of the oligomer size distribution.  For Ab40, the steady-state value of the average length, $<L>$, is about 41.6 for N = 1000 to 40,000 and the peak size of the polymer size distribution is 46 around N =1000 and 51 around N = 5000.  On the other hand, for Ab42-JSS, $<L>$ approaches only about 4.12 for N = 1000 to $<L>$ = 4.43 for N = 10,000 and peak size of the size distribution is 4 for N = 1000, 8 for n = 3000, and 9 for N = 5000. Therefore, the average size of the fibrils or oligomers is much smaller in the case of Ab42-JSS. Taking a closer look at the five sets of parameters, we can attribute the size distribution to two factors: $n_c = n_2 = 3$ for Ab42-JSS and $n_c = n_2 = 2$ for the other 4 sets of parameters and the rate for monomer subtraction is non-zero for Ab42-JSS, but zero for the rest of parameter sets.  Furthermore, it seems that half-life, or the incubation period, is mainly determined by the rate constant for the primary nucleation. The primary nucleation rate of Ab40 is the largest among all five primary nucleation rates. As mentioned above, the values of N and the amount of crowders present may cause the change of $t_{1/2}$ as well.  For higher $\phi$ value, for example. $\phi = 0.2$, $t_{1/2}$ of Ab42-JSS becomes smaller than that of Ab40. 

Next, we return to the question whether we can justify the validity of the Ab42-JSS set of rate parameters in relation to the Ab40 set. According to the current version of the amyloid hypothesis the small, soluble oligomers of $\beta$-amyloid peptides may be more neurotoxic than larger fibrils, such as amyloid plaques. The presence of the high population of the smaller oligomers, such as tetramers to octamers, in the steady-state in the Ab42-JSS set that we mentioned earlier would suggest the higher toxicity of Ab42 compared to Ab40.  This certainly lends support to the validity of Ab42-JSS as a useful set of parameters for A$\beta$42. 

As a side note, at a higher $\phi$ value the smaller $t_{1/2}$ values in the Ab42-JSS set than those of the Ab40 set are observed.  This fact may again help justifying the Ab42-JSS set. In the brain $\beta$-amyloid peptides are present in the space outside the brain cells. This space is confined and distorted, if not crowded.  The confinement effects are to a certain extent similar to the effects of molecular crowding. In this way, the lower values of half-life at higher $\phi$ may to a certain extent lend support to the Ab42-JSS set of rate parameters.

\subsection{Ab40 versus Ab42-Cohen}

\begin{figure}[ht]
    \centering
    \includegraphics[width=\columnwidth]{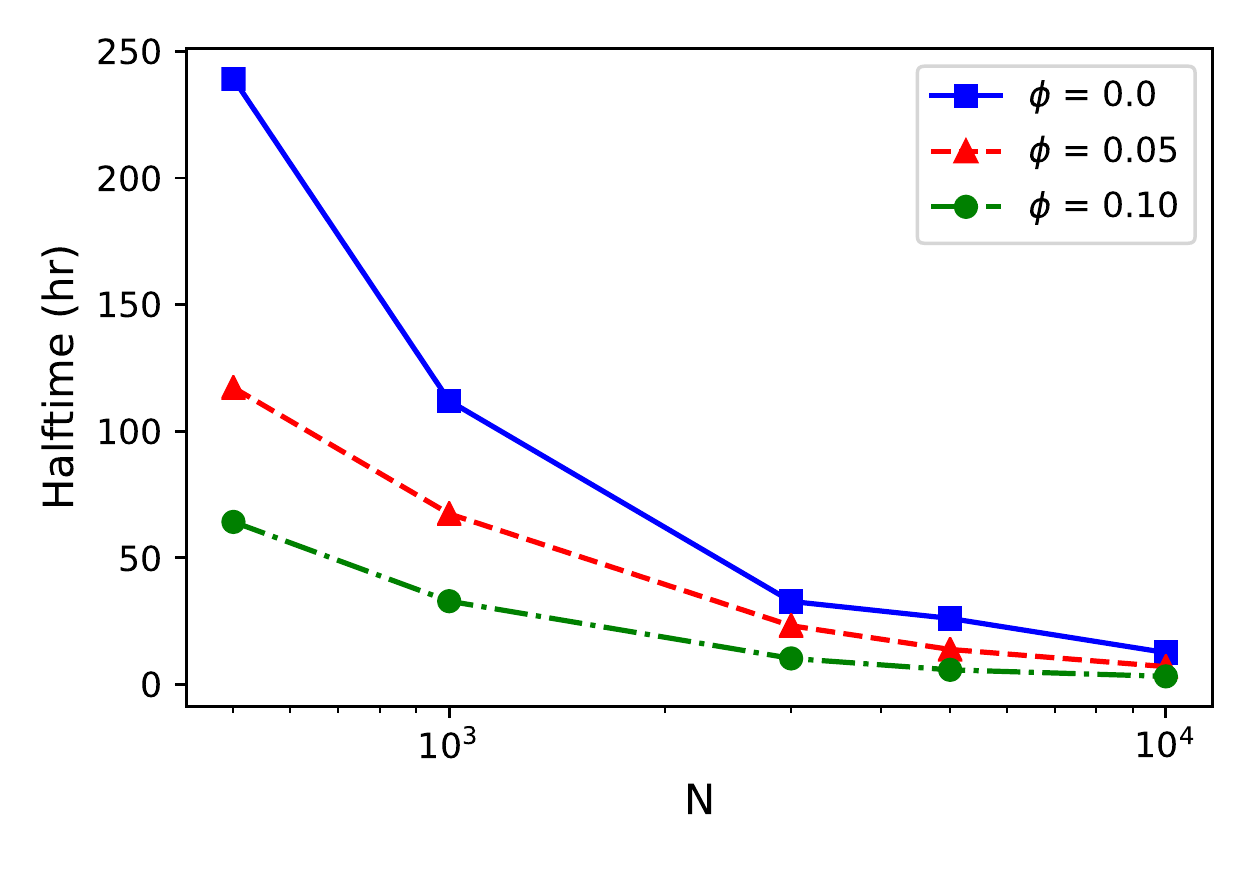}
    \caption{The halftime as a function of N and $\phi$ for the Ab42-Cohen set.}
    \label{fig3}
\end{figure}

As discussed before the primary nucleation rate constant of Ab40 is greater than those of any other set.
For Ab42-Cohen, that rate is about one order of magnitude smaller, thus the halftimes are larger than
those of Ab40 for N ranges from 500 to 10,000. For the range of $\phi$ investigated, from 0 to 0.2, halftimes of Ab42-Cohen stay lager than those of Ab40. For details, we refer to Fig. 3, where the halftime is plotted as a function of N and $\phi$. Incidentally, for these two sets of rate parameters, $n_c$ and $n_2$ are both equal to 2. As in the case of Ab42-JSS, we should check other features of reaction kinetics, such as average length of the fibrils. This does not help either, for the steady-state $<L>$ value is about 56 for Ab42-Cohen and is about 41.6 for Ab40. The peak size of the fibril size distribution for Ab42-Cohen is about 61, compared to 46 for Ab40. Based on the above observations we may suspect the usefulness of the Ab42-Cohen parameter set for the peptide A$\beta$42. However, a chance of saving this set of parameters is at high N values. Using the \textit{popsim} simulator, for N = 40,000 we obtain $t_{1/2} = 3.6$ hr using the Ab42-Cohen parameter set, compared to 9.3 hr for the Ab40 set.
This arises from the fact that as N increases the halftime drops down faster for the Ab42-Cohen set than for the Ab40 set.

\subsection{Ab40-Meisl versus Ab42-Meisl}

\begin{table*}[t]
\begin{center}
\begin{tabular}{ c | c | c | c | c | c }
 & Ab40 &	Ab42-JSS &	Ab42-Cohen &	Ab42-Meisl &	Ab40-Meisl \\
 \hline
$n_c$	& 2	& 3	& 2	& 2	& 2 \\
$n_2$	& 2	& 3	& 2	& 2	& 2 \\
$C_0$ &	\SI{5.0}{\micro \Molar} & \SI{5.0}{\micro \Molar} & \SI{5.0}{\micro \Molar} & \SI{3.5}{\micro \Molar} & \SI{3.5}{\micro \Molar} \\
$a^0$	& \SI{3.6}{\per\micro\Molar\per\hour} &	\SI{8.34}{\per\micro\Molar\per\hour} & \SI{85.0}{\per\micro\Molar\per\hour} &\SI{10800}{\per\micro\Molar\per\hour} & \SI{1080}{\per\micro\Molar\per\hour} \\
$b^0$	& 0	& 0	& 0	& 0	& 0 \\
$k^0_+$	& 0	& 0	& 0 & 0 & 0 \\
$k^0_-$	& 0 & \SI{1.43e-2}{\per\hour} & 0 &	0 &	0 \\
$k^0_n$	& \SI{1.19e-5}{\per\micro\Molar\per\hour} & \SI{9.91e-8}{\per\micro\Molar^{-2}} h$^{-1}$ & \SI{1.72e-6}{\per\micro\Molar\per\hour}	& \SI{1.08e-6}{\per\micro\Molar\per\hour} & \SI{7.2e-9}{\per\micro\Molar\per\hour} \\
$\bar{k}^0_2$	& \SI{6.30e-2}{\per\hour} & \SI{1.74e3}{\per\hour}& \SI{0.276}{\per\hour} & 0 & 0		\\
$k_f$ &	\SI{0.63}{\per\micro\Molar\per\hour} &	\SI{1.72e2}{\per\micro\Molar\per\hour} & \SI{0.0899}{\per\micro\Molar\per\hour} & 0 & 0		\\
$k_b$ &	\SI{30.38}{\per\hour}  & \SI{1.78e9}{\per\hour} & \SI{1.24}{\per\hour} & 0 & 0		\\
$k^0_2$	 & \SI{1.304e-3}{}	& \SI{1.68e-4}{} & 0.0164 & \SI{3.6e-5}{} & \SI{1.08e-5}{} \\
$r^0_c$	 & 1.8 nm &	1.8 nm & 1.8 nm	& 1.8 nm &	1.8 nm \\
$r_1$ &	2.26 nm	& 2.26 nm &	2.26 nm	& 2.26 nm	& 2.26 nm \\
$r_{sc}$ &	2.4 $r_1$ &	2.5 $r_1$	& 2.0 $r_1$	& 2.0 $r_1$	& 2.0 $r_1$ \\
$A_1$ &	\SI{2.35e-8}{\nano\metre^{-xi}} &	\SI{2.35e-8}{\nano\metre^{-xi}} & \SI{2.35e-8}{\nano\metre^{-xi}} & \SI{2.35e-8}{\nano\metre^{-xi}} &	\SI{2.35e-8}{\nano\metre^{-xi}} \\
$xi$ &	0.67 &	0.67 &	0.67 &	0.67 &	0.67 \\
\hline\hline
\end{tabular}
\end{center}
\caption{All 5 sets of reaction rate, length and other parameters. All quantities with superscript 0 denote rates for the crowderless cases. The unit for $k_2$ is $\mu M^{-n_2}$$h^{-1}$.}
\label{table1}
\end{table*}

\begin{table}[t]
\begin{center}
\def\arraystretch{1.5}
\begin{tabular}{ c | c  c  c | c  c  c }
	& \multicolumn{3}{c |}{$N=3000$} & \multicolumn{3}{c }{$N=5000$} \\ 	
	\hline 
$\phi$	& 0	& 0.05 & 0.1 & 0 & 0.05 & 0.1 \\
\hline 
$t_{1/2}$	& 9364 &	4733 &	2263 &	5235 &	3261 &	1354 \\
$\langle L \rangle$	& 2920 &	2690 &	1890 &	4640 &	3730 &	1900 \\
Peak Size	& 2236 & 2087 & 1938 & 3736 & 3487 & 3238 \\
P & 	1 & 1.21 & 1.89 & 1.14 & 1.56 & 2.89 \\ 
\hline\hline
\end{tabular}
\end{center}
\caption{Results for Ab40-Meisl. $t_{1/2}$ halftime; $<L>$, peak size, and P are steady-state values of aggregate length, peak size of the size distribution, and number of aggregates.}
\label{table2}
\end{table}

\begin{table}[t]
\begin{center}
\def\arraystretch{1.5}
\begin{tabular}{ c | c  c  c | c  c  c }
	& \multicolumn{3}{c |}{$N=3000$} & \multicolumn{3}{c }{$N=10000$} \\ 	
	\hline 
$\phi$	& 0	& 0.05 & 0.1 & 0 & 0.05 & 0.1 \\
\hline 
$t_{1/2}$	& 61.29 & 35.25 & 16.44 & 19.95 & 7.69 & 4.53 \\
$\langle L \rangle$	& 2970 & 2890 & 2510 & 9170 & 7280 & 3180 \\
Peak Size	& 2087 & 2236 & 2236 & 7486 & 7486 & 5490 \\
P & 	1.02 & 1.08 & 1.35 & 1.17 & 1.62 & 3.46 \\ 
\hline\hline
\end{tabular}
\end{center}
\caption{Results for Ab42-Meisl.}
\label{table3}
\end{table}

In this subsection, we examine results obtained using \textit{popsim} to simulate the kinetic behaviors of the Ab40-Meisl and Ab42-Meisl sets of parameters and compare their behaviors against those of the other sets of parameters. To start with, both the sizes of the primary nucleus and secondary nucleus are 2 for the Ab40-Meisl and Ab42-Meisl sets. First, from Table I we see that the reaction rate of monomer addition of Ab40-Meissl is one order of magnitude greater than that of Ab42-Cohen and more than two orders of magnitude greater than those of Ab42-JSS and Ab40. The monomer addition rate of Ab42-Meisl is another order of magnitude greater than that of Ab40-Meisl.  On the other hand, the rate of primary nucleation of Ab40-Meisl is the lowest among all sets of parameters, that of Ab42-Meisl is two orders of magnitude higher, which is in the middle of the range. Since the halftime is mainly determined by the rate of primary nucleation, its value for Ab40-Meisl shown in Table II is 9364 hr, or 390 days, longest among all sets of parameters, for N = 3000 and that for Ab42-Meisl as shown in Table III is 61 hr, or 2.5 days. The very large rate of monomer addition means that once the primary nucleus is formed, it will grow long very quick. This yields quite different result from the previous three sets of rate parameters. The stochastic kinetic simulations using \textit{popsim} show in Tables II and III that there is essentially one fibril growing to about the full length. In these tables, $<L>$ denotes the  experimental data value of the average fibril length and P the average number of fibrils.  In the presence of crowders, as $\phi$ increases the chance of creating additional fibrils increases. A similar effect is seen as N increases. However, the number of fibrils remain small. As shown in Table III, the highest average fibril number obtained is 3.46, when $\phi$ increases to 0.1. In general, as the volume fraction of crowers increases, the secondary nucleation becomes more important  The number of aggregates, P, increases and the average length of aggregates, $<L>$. and the size of the peak of the aggregate size distribution, decrease.  As $\phi$ increases above 0.15, secondary nucleation may even overtake monomer addition as the most important reaction mechanism. Due to the overwhelmingly large magnitude, monomer addition is usually the dominating reaction mechanism for the A$\beta$ systems in, at least, the crowderless condition.  

The halftimes shown in Table II for Ab40-Meisl and Table III for Ab42-Meisl seem to support the idea that these sets of rate parameters are useful for the simulations for A$\beta$40 and A$\beta$42, because the rate of growth rate of A$\beta$42 would be much higher than that of A$\beta$40. However, a word of caution is that these sets of rate parameters were obtained through fitting experimental data by solving rate equations \cite{meisl14}. The rate equations, which described how concentrations of chemical species evolve, are effective only when concentrations of chemical species are well-defined.  A concentration, however, is a thermodynamic quantity, which is useful, only when the number of the chemical species involved is large, for the fluctuation is inversely proportional to the number of the chemical species.  The rate equation approach should not be considered justified, if throughout a reaction process the numbers of chemical species remain close to unity.

\section{Comments}
\subsection{System volume}
Stochastic simulations inherently deal with small numbers of particles and, thus, small volumes. It is relatively simple to calculate the volume of the system in question from initial conditions by rearranging the concentration relationship to find
\begin{equation*}
    V = \frac{N_0}{N_Ac_0}.
\end{equation*}
Typically the smaller the simulation volume, the larger fluctuations will be relative to total number of particles. This is where \textit{popsim} and stochastic simulation become increasingly useful.

\subsection{Size distribution}
In this article we have examined five sets of rate parameters for the kinetic behaviors of A$\beta$40 and A$\beta$42.  It turns out quite diverse results are predicted for the two peptides. For instance, the steady-state average aggregate size predicted (for N = 3000) for A$\beta$42 are, respectively, 4.43, 45.3, 2970
based on the Ab42-JSS, Ab42-Cohen, and Ab42-Meisl sets of parameters. Because of such very different kinetic behaviors predicted, it is, in principle, easy to determine which set is more correct by carrying out one additional experimental measurement, for example, the aggregate size distributions. The tools for such measurements are already available, for example, atomic force microscopy \cite{afm} or cryo-EM methods \cite{yang22}. Measurements of the size distributions should be the best way in the present case to determine which sets of rate parameters are the proper ones to use.  

\subsection{Availability}
\textit{popsim} can be accessed by anyone at \textit{www.popsim.xyz} and is being continuously developed by the authors. 
\section{Acknowledgments}

\bibliography{ab4042}

\begin{thebibliography}{38}%
\makeatletter
\providecommand \@ifxundefined [1]{%
 \@ifx{#1\undefined}
}%
\providecommand \@ifnum [1]{%
 \ifnum #1\expandafter \@firstoftwo
 \else \expandafter \@secondoftwo
 \fi
}%
\providecommand \@ifx [1]{%
 \ifx #1\expandafter \@firstoftwo
 \else \expandafter \@secondoftwo
 \fi
}%
\providecommand \natexlab [1]{#1}%
\providecommand \enquote  [1]{``#1''}%
\providecommand \bibnamefont  [1]{#1}%
\providecommand \bibfnamefont [1]{#1}%
\providecommand \citenamefont [1]{#1}%
\providecommand \href@noop [0]{\@secondoftwo}%
\providecommand \href [0]{\begingroup \@sanitize@url \@href}%
\providecommand \@href[1]{\@@startlink{#1}\@@href}%
\providecommand \@@href[1]{\endgroup#1\@@endlink}%
\providecommand \@sanitize@url [0]{\catcode `\\12\catcode `\$12\catcode
  `\&12\catcode `\#12\catcode `\^12\catcode `\_12\catcode `\%12\relax}%
\providecommand \@@startlink[1]{}%
\providecommand \@@endlink[0]{}%
\providecommand \url  [0]{\begingroup\@sanitize@url \@url }%
\providecommand \@url [1]{\endgroup\@href {#1}{\urlprefix }}%
\providecommand \urlprefix  [0]{URL }%
\providecommand \Eprint [0]{\href }%
\providecommand \doibase [0]{https://doi.org/}%
\providecommand \selectlanguage [0]{\@gobble}%
\providecommand \bibinfo  [0]{\@secondoftwo}%
\providecommand \bibfield  [0]{\@secondoftwo}%
\providecommand \translation [1]{[#1]}%
\providecommand \BibitemOpen [0]{}%
\providecommand \bibitemStop [0]{}%
\providecommand \bibitemNoStop [0]{.\EOS\space}%
\providecommand \EOS [0]{\spacefactor3000\relax}%
\providecommand \BibitemShut  [1]{\csname bibitem#1\endcsname}%
\let\auto@bib@innerbib\@empty
\bibitem [{\citenamefont {Dobson}(2003)}]{dobson2003protein}%
  \BibitemOpen
  \bibfield  {author} {\bibinfo {author} {\bibfnamefont {C.~M.}\ \bibnamefont
  {Dobson}},\ }\bibfield  {title} {\bibinfo {title} {Protein folding and
  misfolding},\ }\href@noop {} {\bibfield  {journal} {\bibinfo  {journal}
  {Nature}\ }\textbf {\bibinfo {volume} {426}},\ \bibinfo {pages} {884}
  (\bibinfo {year} {2003})}\BibitemShut {NoStop}%
\bibitem [{\citenamefont {{T. Knowles, C. Waudby, G. Devlin, S. Cohen, A.
  Agguzzi, M. Vendruscolo, E. Terentjev, M. Welland, C.
  Dobson}}(2009)}]{knowles-09}%
  \BibitemOpen
  \bibfield  {author} {\bibinfo {author} {\bibnamefont {{T. Knowles, C. Waudby,
  G. Devlin, S. Cohen, A. Agguzzi, M. Vendruscolo, E. Terentjev, M. Welland, C.
  Dobson}}},\ }\bibfield  {title} {\bibinfo {title} {An analytical solution to
  the kinetics of breakable filament assembly},\ }\href@noop {} {\bibfield
  {journal} {\bibinfo  {journal} {Science}\ }\textbf {\bibinfo {volume}
  {326}},\ \bibinfo {pages} {1533} (\bibinfo {year} {2009})}\BibitemShut
  {NoStop}%
\bibitem [{\citenamefont {Hardy~J}(2002)}]{hardy02amylhypothesis}%
  \BibitemOpen
  \bibfield  {author} {\bibinfo {author} {\bibfnamefont {S.~D.}\ \bibnamefont
  {Hardy~J}},\ }\bibfield  {title} {\bibinfo {title} {The amyloid hypothesis of
  {A}zheimer's disease: progress and problems on the road to therapeutics},\
  }\href@noop {} {\bibfield  {journal} {\bibinfo  {journal} {Science}\ }\textbf
  {\bibinfo {volume} {297}},\ \bibinfo {pages} {353} (\bibinfo {year}
  {2002})}\BibitemShut {NoStop}%
\bibitem [{\citenamefont {Urbanc}(2017)}]{urbanc17}%
  \BibitemOpen
  \bibfield  {author} {\bibinfo {author} {\bibfnamefont {B.}~\bibnamefont
  {Urbanc}},\ }\bibfield  {title} {\bibinfo {title} {Perplexity of amyloid
  $\beta$-protein oligomer formation: Relevance to alzheimer's disease},\ }in\
  \href@noop {} {\emph {\bibinfo {booktitle} {Biophysics and biochemistry of
  protein aggregation}}},\ \bibinfo {editor} {edited by\ \bibinfo {editor}
  {\bibfnamefont {J.~M.}\ \bibnamefont {Yuan}}\ and\ \bibinfo {editor}
  {\bibfnamefont {H.~X.}\ \bibnamefont {Zhou}}}\ (\bibinfo  {publisher} {World
  Scientific},\ \bibinfo {year} {2017})\ pp.\ \bibinfo {pages}
  {1--50}\BibitemShut {NoStop}%
\bibitem [{\citenamefont {Lewczuk}\ \emph {et~al.}(2017)\citenamefont
  {Lewczuk}, \citenamefont {Matzen}, \citenamefont {Blennow}, \citenamefont
  {Parnetti}, \citenamefont {Molinuevo}, \citenamefont {Eusebi}, \citenamefont
  {Kornhuber}, \citenamefont {Morris},\ and\ \citenamefont {Fagan}}]{lewczuk}%
  \BibitemOpen
  \bibfield  {author} {\bibinfo {author} {\bibfnamefont {P.}~\bibnamefont
  {Lewczuk}}, \bibinfo {author} {\bibfnamefont {A.}~\bibnamefont {Matzen}},
  \bibinfo {author} {\bibfnamefont {K.}~\bibnamefont {Blennow}}, \bibinfo
  {author} {\bibfnamefont {L.}~\bibnamefont {Parnetti}}, \bibinfo {author}
  {\bibfnamefont {J.}~\bibnamefont {Molinuevo}}, \bibinfo {author}
  {\bibfnamefont {P.}~\bibnamefont {Eusebi}}, \bibinfo {author} {\bibfnamefont
  {J.}~\bibnamefont {Kornhuber}}, \bibinfo {author} {\bibfnamefont
  {J.}~\bibnamefont {Morris}},\ and\ \bibinfo {author} {\bibfnamefont
  {A.}~\bibnamefont {Fagan}},\ }\bibfield  {title} {\bibinfo {title}
  {Cerebrospinal fluid ${A}\beta42/40$ corresponds better than ${A}\beta42$ to
  amyloid pet in alzheimer's disease},\ }\href@noop {} {\bibfield  {journal}
  {\bibinfo  {journal} {J Alzheimers Dis.}\ }\textbf {\bibinfo {volume} {55}},\
  \bibinfo {pages} {813} (\bibinfo {year} {2017})}\BibitemShut {NoStop}%
\bibitem [{\citenamefont {Mittal}\ \emph {et~al.}(2015)\citenamefont {Mittal},
  \citenamefont {Chowhan},\ and\ \citenamefont
  {Singh}}]{mittal2015macromolecular}%
  \BibitemOpen
  \bibfield  {author} {\bibinfo {author} {\bibfnamefont {S.}~\bibnamefont
  {Mittal}}, \bibinfo {author} {\bibfnamefont {R.~K.}\ \bibnamefont
  {Chowhan}},\ and\ \bibinfo {author} {\bibfnamefont {L.~R.}\ \bibnamefont
  {Singh}},\ }\bibfield  {title} {\bibinfo {title} {Macromolecular crowding:
  Macromolecules friend or foe},\ }\href@noop {} {\bibfield  {journal}
  {\bibinfo  {journal} {Biochim. Biophys. Acta}\ }\textbf {\bibinfo {volume}
  {1850}},\ \bibinfo {pages} {1822} (\bibinfo {year} {2015})}\BibitemShut
  {NoStop}%
\bibitem [{\citenamefont {{R. J. Ellis}}(2001)}]{ellis}%
  \BibitemOpen
  \bibfield  {author} {\bibinfo {author} {\bibnamefont {{R. J. Ellis}}},\
  }\bibfield  {title} {\bibinfo {title} {Macromolecular crowding: Obvious but
  underappreciated},\ }\href@noop {} {\bibfield  {journal} {\bibinfo  {journal}
  {Trends Biochem. Sci.}\ }\textbf {\bibinfo {volume} {26}},\ \bibinfo {pages}
  {597} (\bibinfo {year} {2001})}\BibitemShut {NoStop}%
\bibitem [{\citenamefont {Minton}(1998)}]{Minton98}%
  \BibitemOpen
  \bibfield  {author} {\bibinfo {author} {\bibfnamefont {A.~P.}\ \bibnamefont
  {Minton}},\ }\bibfield  {title} {\bibinfo {title} {Molecular crowding:
  Analysis of effects of high concentrations of inert cosolutes on biochemical
  equilibria and rates in terms of volume exclusion},\ }in\ \href@noop {}
  {\emph {\bibinfo {booktitle} {Methods in enzymology}}},\ \bibinfo {editor}
  {edited by\ \bibinfo {editor} {\bibfnamefont {M.~L.}\ \bibnamefont
  {Johnson}}}\ (\bibinfo  {publisher} {Academic Press},\ \bibinfo {year}
  {1998})\ pp.\ \bibinfo {pages} {127--149}\BibitemShut {NoStop}%
\bibitem [{\citenamefont {Minton}(2014)}]{Minton-14}%
  \BibitemOpen
  \bibfield  {author} {\bibinfo {author} {\bibfnamefont {A.}~\bibnamefont
  {Minton}},\ }\bibfield  {title} {\bibinfo {title} {The effect of
  time-dependent macromolecular crowding on the kinetics of protein
  aggregation: {A} simple model for the onset of age-related neurodegenerative
  disease},\ }\href@noop {} {\bibfield  {journal} {\bibinfo  {journal} {Front.
  Phys.}\ }\textbf {\bibinfo {volume} {12}} (\bibinfo {year}
  {2014})}\BibitemShut {NoStop}%
\bibitem [{\citenamefont {{G. ping, G. Yang, J. M. Yuan}}(2006)}]{ping06}%
  \BibitemOpen
  \bibfield  {author} {\bibinfo {author} {\bibnamefont {{G. ping, G. Yang, J.
  M. Yuan}}},\ }\bibfield  {title} {\bibinfo {title} {Depletion force from
  macromolecular crowding enhances mechanical stability of protein molecules},\
  }\href@noop {} {\bibfield  {journal} {\bibinfo  {journal} {Polymer}\ }\textbf
  {\bibinfo {volume} {47}},\ \bibinfo {pages} {2564} (\bibinfo {year}
  {2006})}\BibitemShut {NoStop}%
\bibitem [{\citenamefont {{J. S. Schreck, J. Bridstrup, and J. M.
  Yuan}}(2017)}]{schreck17}%
  \BibitemOpen
  \bibfield  {author} {\bibinfo {author} {\bibnamefont {{J. S. Schreck, J.
  Bridstrup, and J. M. Yuan}}},\ }\bibfield  {title} {\bibinfo {title} {Kinetic
  studies of protein aggregation with and without the presence of crowders},\
  }in\ \href@noop {} {\emph {\bibinfo {booktitle} {Biophysics and biochemistry
  of protein aggregation}}},\ \bibinfo {editor} {edited by\ \bibinfo {editor}
  {\bibfnamefont {J.~M.}\ \bibnamefont {Yuan}}\ and\ \bibinfo {editor}
  {\bibfnamefont {H.~X.}\ \bibnamefont {Zhou}}}\ (\bibinfo  {publisher} {World
  Scientific},\ \bibinfo {year} {2017})\ pp.\ \bibinfo {pages}
  {187--220}\BibitemShut {NoStop}%
\bibitem [{\citenamefont {Ping}\ \emph {et~al.}(2003)\citenamefont {Ping},
  \citenamefont {Yuan}, \citenamefont {Vallieres}, \citenamefont {Dong},
  \citenamefont {Sun}, \citenamefont {Wei}, \citenamefont {Li},\ and\
  \citenamefont {Lin}}]{ping2003effects}%
  \BibitemOpen
  \bibfield  {author} {\bibinfo {author} {\bibfnamefont {G.}~\bibnamefont
  {Ping}}, \bibinfo {author} {\bibfnamefont {J.}~\bibnamefont {Yuan}}, \bibinfo
  {author} {\bibfnamefont {M.}~\bibnamefont {Vallieres}}, \bibinfo {author}
  {\bibfnamefont {H.}~\bibnamefont {Dong}}, \bibinfo {author} {\bibfnamefont
  {Z.}~\bibnamefont {Sun}}, \bibinfo {author} {\bibfnamefont {Y.}~\bibnamefont
  {Wei}}, \bibinfo {author} {\bibfnamefont {F.}~\bibnamefont {Li}},\ and\
  \bibinfo {author} {\bibfnamefont {S.}~\bibnamefont {Lin}},\ }\bibfield
  {title} {\bibinfo {title} {Effects of confinement on protein folding and
  protein stability},\ }\href@noop {} {\bibfield  {journal} {\bibinfo
  {journal} {The Journal of chemical physics}\ }\textbf {\bibinfo {volume}
  {118}},\ \bibinfo {pages} {8042} (\bibinfo {year} {2003})}\BibitemShut
  {NoStop}%
\bibitem [{\citenamefont {{J. S. Schreck, J. Bridstrup, and J. M.
  Yuan}}(2020)}]{Schreck20}%
  \BibitemOpen
  \bibfield  {author} {\bibinfo {author} {\bibnamefont {{J. S. Schreck, J.
  Bridstrup, and J. M. Yuan}}},\ }\bibfield  {title} {\bibinfo {title}
  {Investigating the effects of molecular crowding on the kinetics of protein
  aggregation},\ }\href@noop {} {\bibfield  {journal} {\bibinfo  {journal} {J.
  Phys. Chem. B.}\ }\textbf {\bibinfo {volume} {124}},\ \bibinfo {pages} {9829}
  (\bibinfo {year} {2020})}\BibitemShut {NoStop}%
\bibitem [{\citenamefont {Nicholson}(2022)}]{nicholson}%
  \BibitemOpen
  \bibfield  {author} {\bibinfo {author} {\bibfnamefont {C.}~\bibnamefont
  {Nicholson}},\ }\bibfield  {title} {\bibinfo {title} {The secret world in the
  gaps between brain cells},\ }\href@noop {} {\bibfield  {journal} {\bibinfo
  {journal} {Physics Today}\ }\textbf {\bibinfo {volume} {75}},\ \bibinfo
  {pages} {26} (\bibinfo {year} {2022})}\BibitemShut {NoStop}%
\bibitem [{\citenamefont {Nicholson}\ and\ \citenamefont
  {Kamali-Zare}(2020)}]{nicholson20}%
  \BibitemOpen
  \bibfield  {author} {\bibinfo {author} {\bibfnamefont {C.}~\bibnamefont
  {Nicholson}}\ and\ \bibinfo {author} {\bibfnamefont {P.}~\bibnamefont
  {Kamali-Zare}},\ }\bibfield  {title} {\bibinfo {title} {Reduction of
  dimensionality in monte carlo simulation of diffusion in extracellular space
  surrounding cubic cells},\ }\href@noop {} {\bibfield  {journal} {\bibinfo
  {journal} {Neurochem. Res.}\ }\textbf {\bibinfo {volume} {45}},\ \bibinfo
  {pages} {42} (\bibinfo {year} {2020})}\BibitemShut {NoStop}%
\bibitem [{\citenamefont {Meisl}\ \emph
  {et~al.}(2016{\natexlab{a}})\citenamefont {Meisl}, \citenamefont {Yang},
  \citenamefont {Frohm}, \citenamefont {Knowles},\ and\ \citenamefont
  {Linse}}]{meisl16}%
  \BibitemOpen
  \bibfield  {author} {\bibinfo {author} {\bibfnamefont {G.}~\bibnamefont
  {Meisl}}, \bibinfo {author} {\bibfnamefont {X.}~\bibnamefont {Yang}},
  \bibinfo {author} {\bibfnamefont {B.}~\bibnamefont {Frohm}}, \bibinfo
  {author} {\bibfnamefont {T.~P.}\ \bibnamefont {Knowles}},\ and\ \bibinfo
  {author} {\bibfnamefont {S.}~\bibnamefont {Linse}},\ }\bibfield  {title}
  {\bibinfo {title} {Quantitative analysis of intrinsic and extrinsic factors
  in the aggregation mechanism of alzheimer-associated a$\beta$-peptide},\
  }\href@noop {} {\bibfield  {journal} {\bibinfo  {journal} {Scientific
  reports}\ }\textbf {\bibinfo {volume} {6}},\ \bibinfo {pages} {18728}
  (\bibinfo {year} {2016}{\natexlab{a}})}\BibitemShut {NoStop}%
\bibitem [{\citenamefont {Meisl}\ \emph
  {et~al.}(2016{\natexlab{b}})\citenamefont {Meisl}, \citenamefont
  {Kirkegaard}, \citenamefont {Arosio}, \citenamefont {Michaels}, \citenamefont
  {Vendruscolo}, \citenamefont {Dobson}, \citenamefont {Linse},\ and\
  \citenamefont {Knowles}}]{meisl2016molecular}%
  \BibitemOpen
  \bibfield  {author} {\bibinfo {author} {\bibfnamefont {G.}~\bibnamefont
  {Meisl}}, \bibinfo {author} {\bibfnamefont {J.~B.}\ \bibnamefont
  {Kirkegaard}}, \bibinfo {author} {\bibfnamefont {P.}~\bibnamefont {Arosio}},
  \bibinfo {author} {\bibfnamefont {T.~C.}\ \bibnamefont {Michaels}}, \bibinfo
  {author} {\bibfnamefont {M.}~\bibnamefont {Vendruscolo}}, \bibinfo {author}
  {\bibfnamefont {C.~M.}\ \bibnamefont {Dobson}}, \bibinfo {author}
  {\bibfnamefont {S.}~\bibnamefont {Linse}},\ and\ \bibinfo {author}
  {\bibfnamefont {T.~P.}\ \bibnamefont {Knowles}},\ }\bibfield  {title}
  {\bibinfo {title} {Molecular mechanisms of protein aggregation from global
  fitting of kinetic models},\ }\href@noop {} {\bibfield  {journal} {\bibinfo
  {journal} {Nat. Protoc.}\ }\textbf {\bibinfo {volume} {11}},\ \bibinfo
  {pages} {252} (\bibinfo {year} {2016}{\natexlab{b}})}\BibitemShut {NoStop}%
\bibitem [{\citenamefont {{G. Meisl, X. Yang, E. Hellstrand, B. Frohm, J. B.
  Kirkegraard, S. I. A. Cohen, C. M. Dobson, S. Linse, T. P. J.
  Knowles}}(2014)}]{meisl14}%
  \BibitemOpen
  \bibfield  {author} {\bibinfo {author} {\bibnamefont {{G. Meisl, X. Yang, E.
  Hellstrand, B. Frohm, J. B. Kirkegraard, S. I. A. Cohen, C. M. Dobson, S.
  Linse, T. P. J. Knowles}}},\ }\bibfield  {title} {\bibinfo {title}
  {Differences in nucleation behavior underlie the contrasting aggregation
  kinetics of the a$\beta$40 and a$\beta$42 peptides},\ }\href@noop {}
  {\bibfield  {journal} {\bibinfo  {journal} {Proc. Natl. Acad. Sci. U.S.A.}\
  }\textbf {\bibinfo {volume} {111}},\ \bibinfo {pages} {9384} (\bibinfo {year}
  {2014})}\BibitemShut {NoStop}%
\bibitem [{\citenamefont {Cohen}\ \emph {et~al.}(2013)\citenamefont {Cohen},
  \citenamefont {Linse}, \citenamefont {Luheshi}, \citenamefont {Hellstrand},
  \citenamefont {White}, \citenamefont {Rajah}, \citenamefont {Otzen},
  \citenamefont {Vendruscolo}, \citenamefont {Dobson},\ and\ \citenamefont
  {Knowles}}]{cohen2013proliferation}%
  \BibitemOpen
  \bibfield  {author} {\bibinfo {author} {\bibfnamefont {S.~I.}\ \bibnamefont
  {Cohen}}, \bibinfo {author} {\bibfnamefont {S.}~\bibnamefont {Linse}},
  \bibinfo {author} {\bibfnamefont {L.~M.}\ \bibnamefont {Luheshi}}, \bibinfo
  {author} {\bibfnamefont {E.}~\bibnamefont {Hellstrand}}, \bibinfo {author}
  {\bibfnamefont {D.~A.}\ \bibnamefont {White}}, \bibinfo {author}
  {\bibfnamefont {L.}~\bibnamefont {Rajah}}, \bibinfo {author} {\bibfnamefont
  {D.~E.}\ \bibnamefont {Otzen}}, \bibinfo {author} {\bibfnamefont
  {M.}~\bibnamefont {Vendruscolo}}, \bibinfo {author} {\bibfnamefont {C.~M.}\
  \bibnamefont {Dobson}},\ and\ \bibinfo {author} {\bibfnamefont {T.~P.}\
  \bibnamefont {Knowles}},\ }\bibfield  {title} {\bibinfo {title}
  {Proliferation of amyloid-$\beta$42 aggregates occurs through a secondary
  nucleation mechanism},\ }\href@noop {} {\bibfield  {journal} {\bibinfo
  {journal} {Proc. Natl. Acad. Sci. U.S.A.}\ }\textbf {\bibinfo {volume}
  {110}},\ \bibinfo {pages} {9758} (\bibinfo {year} {2013})}\BibitemShut
  {NoStop}%
\bibitem [{\citenamefont {Szavits-Nossan}\ \emph {et~al.}(2014)\citenamefont
  {Szavits-Nossan}, \citenamefont {Eden}, \citenamefont {Morris}, \citenamefont
  {MacPhee}, \citenamefont {Evans},\ and\ \citenamefont {Allen}}]{allen}%
  \BibitemOpen
  \bibfield  {author} {\bibinfo {author} {\bibfnamefont {J.}~\bibnamefont
  {Szavits-Nossan}}, \bibinfo {author} {\bibfnamefont {K.}~\bibnamefont
  {Eden}}, \bibinfo {author} {\bibfnamefont {R.~J.}\ \bibnamefont {Morris}},
  \bibinfo {author} {\bibfnamefont {C.~E.}\ \bibnamefont {MacPhee}}, \bibinfo
  {author} {\bibfnamefont {M.~R.}\ \bibnamefont {Evans}},\ and\ \bibinfo
  {author} {\bibfnamefont {R.~J.}\ \bibnamefont {Allen}},\ }\bibfield  {title}
  {\bibinfo {title} {Inherent variability in the kinetics of autocatalytic
  protein self-assembly},\ }\href
  {https://doi.org/10.1103/PhysRevLett.113.098101} {\bibfield  {journal}
  {\bibinfo  {journal} {Phys. Rev. Lett.}\ }\textbf {\bibinfo {volume} {113}},\
  \bibinfo {pages} {098101} (\bibinfo {year} {2014})}\BibitemShut {NoStop}%
\bibitem [{\citenamefont {Tiwari}\ and\ \citenamefont {van~der
  Schoot}(2016)}]{tiwari}%
  \BibitemOpen
  \bibfield  {author} {\bibinfo {author} {\bibfnamefont {N.~S.}\ \bibnamefont
  {Tiwari}}\ and\ \bibinfo {author} {\bibfnamefont {P.}~\bibnamefont {van~der
  Schoot}},\ }\bibfield  {title} {\bibinfo {title} {Stochastic lag time in
  nucleated linear self-assembly},\ }\href@noop {} {\bibfield  {journal}
  {\bibinfo  {journal} {J. Chem. Phys.}\ }\textbf {\bibinfo {volume} {144}},\
  \bibinfo {pages} {235101} (\bibinfo {year} {2016})}\BibitemShut {NoStop}%
\bibitem [{\citenamefont {{J. Bridstrup, J. S. Schreck, and J. M.
  Yuan}}(2021)}]{bridstrup21}%
  \BibitemOpen
  \bibfield  {author} {\bibinfo {author} {\bibnamefont {{J. Bridstrup, J. S.
  Schreck, and J. M. Yuan}}},\ }\bibfield  {title} {\bibinfo {title}
  {Stochastic kinetic treatment of protein aggregation and the effects of
  macromolecular crowding},\ }\href@noop {} {\bibfield  {journal} {\bibinfo
  {journal} {J. Phys. Chem. B.}\ }\textbf {\bibinfo {volume} {125}},\ \bibinfo
  {pages} {6068} (\bibinfo {year} {2021})}\BibitemShut {NoStop}%
\bibitem [{\citenamefont {Shen}\ \emph {et~al.}(2021)\citenamefont {Shen},
  \citenamefont {Tsai}, \citenamefont {Schafer},\ and\ \citenamefont
  {Wolynes}}]{shen21stochastic}%
  \BibitemOpen
  \bibfield  {author} {\bibinfo {author} {\bibfnamefont {J.~L.}\ \bibnamefont
  {Shen}}, \bibinfo {author} {\bibfnamefont {M.~Y.}\ \bibnamefont {Tsai}},
  \bibinfo {author} {\bibfnamefont {N.~P.}\ \bibnamefont {Schafer}},\ and\
  \bibinfo {author} {\bibfnamefont {P.~G.}\ \bibnamefont {Wolynes}},\
  }\bibfield  {title} {\bibinfo {title} {Modeling protein aggregation kinetics:
  The method of second stochasticization},\ }\href@noop {} {\bibfield
  {journal} {\bibinfo  {journal} {J. Phys. Chem. B}\ }\textbf {\bibinfo
  {volume} {125}},\ \bibinfo {pages} {1118} (\bibinfo {year}
  {2021})}\BibitemShut {NoStop}%
\bibitem [{\citenamefont {Gillespie}(2007)}]{gillespie07}%
  \BibitemOpen
  \bibfield  {author} {\bibinfo {author} {\bibfnamefont {D.~T.}\ \bibnamefont
  {Gillespie}},\ }\bibfield  {title} {\bibinfo {title} {Stochastic simulation
  of chemical kinetics},\ }\href@noop {} {\bibfield  {journal} {\bibinfo
  {journal} {Annu. Rev. Phys. Chem.}\ }\textbf {\bibinfo {volume} {58}},\
  \bibinfo {pages} {35–55} (\bibinfo {year} {2007})}\BibitemShut {NoStop}%
\bibitem [{\citenamefont {Gillespie}(1992)}]{gillespie92}%
  \BibitemOpen
  \bibfield  {author} {\bibinfo {author} {\bibfnamefont {D.~T.}\ \bibnamefont
  {Gillespie}},\ }\bibfield  {title} {\bibinfo {title} {A rigorous derivation
  of the chemical master equation},\ }\href@noop {} {\bibfield  {journal}
  {\bibinfo  {journal} {Physica A}\ }\textbf {\bibinfo {volume} {188}},\
  \bibinfo {pages} {404} (\bibinfo {year} {1992})}\BibitemShut {NoStop}%
\bibitem [{\citenamefont {Gillespie}(1977)}]{gillespie77}%
  \BibitemOpen
  \bibfield  {author} {\bibinfo {author} {\bibfnamefont {D.~T.}\ \bibnamefont
  {Gillespie}},\ }\bibfield  {title} {\bibinfo {title} {Exact stochastic
  simulation of coupled chemical reactions},\ }\href@noop {} {\bibfield
  {journal} {\bibinfo  {journal} {J. Phys. Chem.}\ }\textbf {\bibinfo {volume}
  {81}},\ \bibinfo {pages} {2340–61} (\bibinfo {year} {1977})}\BibitemShut
  {NoStop}%
\bibitem [{\citenamefont {Bridstrup}\ and\ \citenamefont
  {Yuan}(2016)}]{bridstrup16}%
  \BibitemOpen
  \bibfield  {author} {\bibinfo {author} {\bibfnamefont {J.}~\bibnamefont
  {Bridstrup}}\ and\ \bibinfo {author} {\bibfnamefont {J.~M.}\ \bibnamefont
  {Yuan}},\ }\bibfield  {title} {\bibinfo {title} {Effects of crowders on the
  equilibrium and kinetic properties of protein aggregation},\ }\href@noop {}
  {\bibfield  {journal} {\bibinfo  {journal} {Chem. Phys. Lett.}\ }\textbf
  {\bibinfo {volume} {659}},\ \bibinfo {pages} {252} (\bibinfo {year}
  {2016})}\BibitemShut {NoStop}%
\bibitem [{\citenamefont {Becker}\ and\ \citenamefont {Doring}(1935)}]{BD}%
  \BibitemOpen
  \bibfield  {author} {\bibinfo {author} {\bibfnamefont {R.}~\bibnamefont
  {Becker}}\ and\ \bibinfo {author} {\bibfnamefont {W.}~\bibnamefont
  {Doring}},\ }\bibfield  {title} {\bibinfo {title} {Kinetische behandlung der
  keimbildung in ubersattigten dampfern},\ }\href@noop {} {\bibfield  {journal}
  {\bibinfo  {journal} {Ann. Phys}\ }\textbf {\bibinfo {volume} {24}},\
  \bibinfo {pages} {719} (\bibinfo {year} {1935})}\BibitemShut {NoStop}%
\bibitem [{\citenamefont {Wattis}(2006)}]{Wattis}%
  \BibitemOpen
  \bibfield  {author} {\bibinfo {author} {\bibfnamefont {J.}~\bibnamefont
  {Wattis}},\ }\bibfield  {title} {\bibinfo {title} {An introduction to
  mathematical models of coagulation-fragmentation processes: A discrete
  deterministic mean-field approach},\ }\href@noop {} {\bibfield  {journal}
  {\bibinfo  {journal} {Physica D}\ }\textbf {\bibinfo {volume} {222}},\
  \bibinfo {pages} {1} (\bibinfo {year} {2006})}\BibitemShut {NoStop}%
\bibitem [{\citenamefont {Oosawa}\ and\ \citenamefont
  {Asakura}(1975)}]{oosawa}%
  \BibitemOpen
  \bibfield  {author} {\bibinfo {author} {\bibfnamefont {F.}~\bibnamefont
  {Oosawa}}\ and\ \bibinfo {author} {\bibfnamefont {S.}~\bibnamefont
  {Asakura}},\ }\href@noop {} {\emph {\bibinfo {title} {Thermodynamics of the
  Polymerization of Protein}}}\ (\bibinfo  {publisher} {Academic Press},\
  \bibinfo {year} {1975})\BibitemShut {NoStop}%
\bibitem [{\citenamefont {Schreck}\ and\ \citenamefont
  {Yuan}(2013)}]{Schreck13}%
  \BibitemOpen
  \bibfield  {author} {\bibinfo {author} {\bibfnamefont {J.~S.}\ \bibnamefont
  {Schreck}}\ and\ \bibinfo {author} {\bibfnamefont {J.~M.}\ \bibnamefont
  {Yuan}},\ }\bibfield  {title} {\bibinfo {title} {A kinetic study of amyloid
  formation: Fibril growth and length distributions},\ }\href@noop {}
  {\bibfield  {journal} {\bibinfo  {journal} {J. Phys. Chem. B.}\ }\textbf
  {\bibinfo {volume} {117}},\ \bibinfo {pages} {6574} (\bibinfo {year}
  {2013})}\BibitemShut {NoStop}%
\bibitem [{\citenamefont {Wattis}(1999)}]{wattis1999becker}%
  \BibitemOpen
  \bibfield  {author} {\bibinfo {author} {\bibfnamefont {J.~A.}\ \bibnamefont
  {Wattis}},\ }\bibfield  {title} {\bibinfo {title} {A becker-d{\"o}ring model
  of competitive nucleation},\ }\href@noop {} {\bibfield  {journal} {\bibinfo
  {journal} {J. Phys. A}\ }\textbf {\bibinfo {volume} {32}},\ \bibinfo {pages}
  {8755} (\bibinfo {year} {1999})}\BibitemShut {NoStop}%
\bibitem [{\citenamefont {{H. Reiss, H. L. Frisch, J. L.
  Lebowitz}}(1959)}]{SPT}%
  \BibitemOpen
  \bibfield  {author} {\bibinfo {author} {\bibnamefont {{H. Reiss, H. L.
  Frisch, J. L. Lebowitz}}},\ }\bibfield  {title} {\bibinfo {title}
  {Statistical mechanics of rigid spheres},\ }\href@noop {} {\bibfield
  {journal} {\bibinfo  {journal} {J. Chem. Phys.}\ }\textbf {\bibinfo {volume}
  {31}},\ \bibinfo {pages} {369} (\bibinfo {year} {1959})}\BibitemShut
  {NoStop}%
\bibitem [{\citenamefont {{D. Hall and A.P. Minton}}(2002)}]{hall02}%
  \BibitemOpen
  \bibfield  {author} {\bibinfo {author} {\bibnamefont {{D. Hall and A.P.
  Minton}}},\ }\bibfield  {title} {\bibinfo {title} {Effects of inert
  volume-excluding macromolecules on protein fiber formation. {I}.
  {E}quilibrium models},\ }\href@noop {} {\bibfield  {journal} {\bibinfo
  {journal} {Biophys. Chem.}\ }\textbf {\bibinfo {volume} {98}},\ \bibinfo
  {pages} {93} (\bibinfo {year} {2002})}\BibitemShut {NoStop}%
\bibitem [{\citenamefont {{D. Hall and A.P. Minton}}(2004)}]{hall04}%
  \BibitemOpen
  \bibfield  {author} {\bibinfo {author} {\bibnamefont {{D. Hall and A.P.
  Minton}}},\ }\bibfield  {title} {\bibinfo {title} {Effects of inert
  volume-excluding macromolecules on protein fiber formation. {II}. {K}inetic
  models for nucleated fiber growth},\ }\href@noop {} {\bibfield  {journal}
  {\bibinfo  {journal} {Biophys. Chem.}\ }\textbf {\bibinfo {volume} {107}},\
  \bibinfo {pages} {299} (\bibinfo {year} {2004})}\BibitemShut {NoStop}%
\bibitem [{\citenamefont {{F. A. Ferrone, M. Ivanova, R.
  Jasuja}}(2002)}]{Ferrone02}%
  \BibitemOpen
  \bibfield  {author} {\bibinfo {author} {\bibnamefont {{F. A. Ferrone, M.
  Ivanova, R. Jasuja}}},\ }\bibfield  {title} {\bibinfo {title} {Heterogeneous
  nucleation and crowding in sickle hemoglobin: An analytic approach},\
  }\href@noop {} {\bibfield  {journal} {\bibinfo  {journal} {Biophys. J.}\
  }\textbf {\bibinfo {volume} {82}},\ \bibinfo {pages} {399} (\bibinfo {year}
  {2002})}\BibitemShut {NoStop}%
\bibitem [{\citenamefont {Ruggeri}\ \emph {et~al.}(2019)\citenamefont
  {Ruggeri}, \citenamefont {Šneideris}, \citenamefont {Vendruscolo},\ and\
  \citenamefont {Knowles}}]{afm}%
  \BibitemOpen
  \bibfield  {author} {\bibinfo {author} {\bibfnamefont {F.}~\bibnamefont
  {Ruggeri}}, \bibinfo {author} {\bibfnamefont {T.}~\bibnamefont {Šneideris}},
  \bibinfo {author} {\bibfnamefont {M.}~\bibnamefont {Vendruscolo}},\ and\
  \bibinfo {author} {\bibfnamefont {T.}~\bibnamefont {Knowles}},\ }\bibfield
  {title} {\bibinfo {title} {Atomic force microscopy for single molecule
  characterisation of protein aggregation},\ }\href@noop {} {\bibfield
  {journal} {\bibinfo  {journal} {Arch Biochem Biophys.}\ }\textbf {\bibinfo
  {volume} {664}},\ \bibinfo {pages} {134} (\bibinfo {year}
  {2019})}\BibitemShut {NoStop}%
\bibitem [{\citenamefont {Yang}\ \emph {et~al.}(2022)\citenamefont {Yang},
  \citenamefont {Arseni}, \citenamefont {Zhang}, \citenamefont {Huang},
  \citenamefont {Lövestam}, \citenamefont {Schweighauser}, \citenamefont
  {Kotecha}, \citenamefont {Murzin}, \citenamefont {Peak-Chew}, \citenamefont
  {Macdonald}, \citenamefont {Lavenir}, \citenamefont {Garringer},
  \citenamefont {Gelpi}, \citenamefont {Newell}, \citenamefont {Kovacs},
  \citenamefont {Vidal}, \citenamefont {Ghetti}, \citenamefont
  {Ryskeldi-Falcon}, \citenamefont {Scheres},\ and\ \citenamefont
  {Goedert}}]{yang22}%
  \BibitemOpen
  \bibfield  {author} {\bibinfo {author} {\bibfnamefont {Y.}~\bibnamefont
  {Yang}}, \bibinfo {author} {\bibfnamefont {D.}~\bibnamefont {Arseni}},
  \bibinfo {author} {\bibfnamefont {W.}~\bibnamefont {Zhang}}, \bibinfo
  {author} {\bibfnamefont {M.}~\bibnamefont {Huang}}, \bibinfo {author}
  {\bibfnamefont {S.}~\bibnamefont {Lövestam}}, \bibinfo {author}
  {\bibfnamefont {M.}~\bibnamefont {Schweighauser}}, \bibinfo {author}
  {\bibfnamefont {A.}~\bibnamefont {Kotecha}}, \bibinfo {author} {\bibfnamefont
  {A.}~\bibnamefont {Murzin}}, \bibinfo {author} {\bibfnamefont
  {S.}~\bibnamefont {Peak-Chew}}, \bibinfo {author} {\bibfnamefont
  {J.}~\bibnamefont {Macdonald}}, \bibinfo {author} {\bibfnamefont
  {I.}~\bibnamefont {Lavenir}}, \bibinfo {author} {\bibfnamefont
  {H.}~\bibnamefont {Garringer}}, \bibinfo {author} {\bibfnamefont
  {E.}~\bibnamefont {Gelpi}}, \bibinfo {author} {\bibfnamefont
  {K.}~\bibnamefont {Newell}}, \bibinfo {author} {\bibfnamefont
  {G.}~\bibnamefont {Kovacs}}, \bibinfo {author} {\bibfnamefont
  {R.}~\bibnamefont {Vidal}}, \bibinfo {author} {\bibfnamefont
  {B.}~\bibnamefont {Ghetti}}, \bibinfo {author} {\bibfnamefont
  {B.}~\bibnamefont {Ryskeldi-Falcon}}, \bibinfo {author} {\bibfnamefont
  {S.}~\bibnamefont {Scheres}},\ and\ \bibinfo {author} {\bibfnamefont
  {M.}~\bibnamefont {Goedert}},\ }\bibfield  {title} {\bibinfo {title} {Cryo-em
  structures of amyloid-$\beta$ 42 filaments from human brains},\ }\href@noop
  {} {\bibfield  {journal} {\bibinfo  {journal} {Science}\ }\textbf {\bibinfo
  {volume} {375}},\ \bibinfo {pages} {167} (\bibinfo {year}
  {2022})}\BibitemShut {NoStop}%
\end{thebibliography}%

\end{document}